\documentclass[prl,twocolumn,showpacs,nofootinbib]{revtex4}
\usepackage[dvips]{graphicx}
\usepackage{color,array,dcolumn}
\usepackage{amsmath}
\usepackage{amssymb}
\usepackage{amsmath}
\usepackage{amssymb}
\begin{document}
\title{ GAUGE APPROACH TO THE SPECIFIC HEAT \\ IN THE NORMAL STATE OF CUPRATES}

\author{P.A. Marchetti}
\affiliation{Dipartimento di Fisica ``G. Galilei'', University of Padova INFN, I-35131
Padova, Italy}
\author{A. Ambrosetti}
\affiliation{Dipartimento di Fisica, INFN, University of Trento I-38100
Povo (TN), Italy}
\date{\today}
 \begin{abstract}

Many experimental features of the electronic specific heat and
entropy of high Tc cuprates in the normal state, including the
nontrivial temperature dependence of the specific heat coefficient
$\gamma$ and negative intercept of the extrapolated entropy to $T=0$
for underdoped cuprates, are reproduced using the spin-charge
gauge approach to
 the $t-J$ model. The entropy turns out to be basically due to fermionic excitations,
 but with a temperature dependence of the specific heat coefficient controlled by
 fluctuations of a  gauge field coupling them to gapful bosonic excitations.
  In particular the negative intercept of the extrapolated entropy at $T=0$ in the
  pseudogap ``phase'' is attributed to the scalar component of the gauge field,
  which implements the  local no-double occupancy constraint.
 \end{abstract}

\maketitle
The low-temperature electronic entropy of high $T_c$ hole-doped cuprates in the
 normal ( ``metallic'') state exhibits a behavior rather unusual for a metal:
 the specific heat coefficient $\gamma$ which should be constant shows a non trivial
 temperature dependence \cite{loram} and even more spectacularly the $T=0$ intercept
 of the entropy, extrapolated from the approximate linear behavior at moderate
 temperatures, turns out to be negative in the underdoped region \cite{exp}.
In this paper we apply the spin-charge gauge approach developed in
\cite{MSY,Mar,Mar1,mar} to extract from the two-dimensional $t$-$J$ model the low
temperature entropy and specific heat in the normal state and compare our results
 with the experimental data, in particular showing how this approach can explain the
 peculiar behavior mentioned above.

Let us first outline the main features of the experiments
\cite{loram, exp, momono}, following Ref. \onlinecite{loram}. In
the normal state of La$_{2-\delta}$Sr$_\delta$CuO$_4$ (LSCO) and YBa$_2$Cu$_3$O$_{6+x}$ (YBCO) the electronic specific heat
coefficient $\gamma=C^{el}/T$ as a function doping concentration
$\delta$ and temperature up
 to 250-300 K exhibits the following behavior.
For strongly underdoped samples $\gamma(\delta,T) \sim \Delta
\gamma^{LT} + B(\delta, T)$ where $\Delta \gamma^{LT}$ is a
Low-Temperature upturn, $ B(\delta, T)$ is slowly increasing
roughly linearly in $\delta$ and $T$ with almost
$\delta$-independent slope. At higher dopings the Low-Temperature
upturn disappears and the
 samples become superconducting. The increasing part remains, but it saturates to a
  broad maximum at $T^*_\gamma$, followed by a slow decrease in $T$. $T^*_\gamma$
  roughly coincides with the pseudogap temperature $T^*$ identified by the inflection
   point in the in-plane resistivity, as can be checked using the data on curvature
   of resistivity \cite{ando}, see Fig 1.
\begin{figure}
\centering
\includegraphics[scale=0.5]{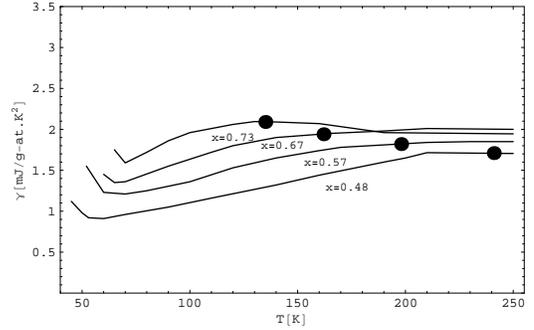}
\caption{The dots mark the inflection point on resistivity (T*) as given in \cite{ando} on $\gamma$ data  in YBa$_2$Cu$_3$O$_{6+x}$ taken from \cite{Bi}; semiempirically $\delta \approx 0.2 x$, see \cite{Bi}.}
\end{figure}
 In the decreasing region $\gamma$ becomes almost
   $\delta$-independent. Similar features are exhibited, in the appropriate doping
   range, also by Bi$_2$Sr$_2$CaCu$_2$O$_{8+x}$ (Bi2212) data \cite{Bi}, so they can thus be considered as rather
   generic in cuprates and therefore it is reasonable to explore a physical
   interpretation in
     terms of the physics of doped Mott insulator described by the $t-J$ model.

 An attempt to discuss the specific heat coefficient in terms of $SU(2)$ slave boson
  theory \cite{lee} of the $t-J$ model appears in Ref \onlinecite{kim}, obtaining for
   the gauge contribution $\gamma \approx -T \ln T$ and some data are fitted with this
   formula. Further attempts to understand the behavior of the specific heat can be
   found in \cite{curty} \cite{varma}.

In spite of the complex features of $\gamma(\delta, T)$ described
above, the entropy $S(\delta,T)$ of the normal state exhibits a
simpler continuity: for low dopings it is
 approximately linear in $T$ with increasing slope as $\delta$ increases, at higher $T$
 the slope becomes almost $\delta$-independent with increasing but negative intercept
 at $T=0$. At higher dopings ($\delta \gtrsim 0.19$) also the intercept becomes almost
  $\delta$-independent and approximately 0. On the basis of the above continuity it was
  argued in Ref. \onlinecite{loram} that in YBCO the spin excitations may be relevant
  for all $\delta$ and the above results are better described as a modification of the
  low-energy spin spectrum as $\delta$ changes than by a simple band model. The
  spin-charge gauge approach appears to partially substantiate such claims. The negative
   intercept of the entropy suggests a negative contribution to entropy of a
   ``constraint-field'' , which acts reducing the low-energy degrees of freedom or
    more precisely removing them from the temperature/energy region considered. This
    was proposed in the analysis of the thermodynamics of the $t-J$ model performed in
     \cite{rice} within the slave-boson approach. In fact, a negative contribution to
     entropy naturally arises in a gauge approach from the scalar component of the
     gauge field (in the Coulomb gauge) enforcing ``Gauss law''.
Let us now sketch the basis of the spin-charge gauge approach and
its application to the computation of entropy and specific heat.

This approach assumes as a (simplified) model for CuO layers in
high $T_c$ cuprates the 2D $t$-$J$ model with $t/J \sim 3$.
Neglecting $t',t''$, details of Fermi surface (FS) are lost but
the analysis is simplified, hopefully retaining the basic relevant
features. The model is treated in an ``improved Mean Field
Approximation''(MFA) via a gauge theory of spin-charge
decomposition, obtained by gauging the global spin and charge
symmetries of the model \cite{Fro}. This gauging is obtained
introducing spin and charge Chern-Simons gauge fields. The nice
feature of introducing these gauge fields is the possibility of a
more flexible treatment of charge and spin responses within a a
spin-charge decomposition scheme. In the end they will disappear
from the game in MFA, but leaving behind sign of their presence
crucial for the low-energy physics, as discussed below. 

The basic fields adopted in this approach for the spin-charge decomposition
of the $t-J$ model are a charged spinless fermion, the holon
\cite{nota}, a neutral spin 1/2 boson of a non-linear $\sigma$
($CP^1$) model, the spinon, and a slave-particle gauge field (not
to be confused with spin and charge Chern-Simons gauge
 fields). The spin-gauge field in MFA attaches spin vortices to the
 empty-site positions. The spinons moving across this gas of vortices
 acquire a mass gap, with a theoretically derived doping dependence,
  $m_s \sim \sqrt{|\delta \ln \delta|}$ consistent with AF correlation
   length at small $\delta$ derived from neutron experiments \cite{Birgenau}.
   In MFA at low temperature and small doping concentration,
 in the parameter region to be compared with the ``pseudogap phase'' (PG) of
  the cuprates, the holons move in a statistical magnetic field with flux $\pi$
  per plaquette generated by the charge-gauge field. This ``phase'' shares some similarity with the $\pi$-flux phase appearing in the slave boson formalism \cite{Lee}. 

Around the pseudogap
  temperature $T^*$  the $\pi$-flux lattice ``melts'' and we enter in the
   ``strange metal phase'' (SM), at higher $\delta$ or $T$ , see Fig. 2. 
\begin{figure}
\centering
\includegraphics[scale=0.37]{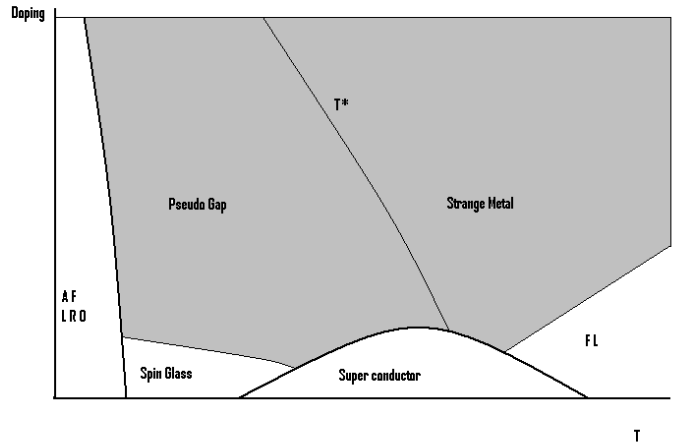}
\caption{Qualitative phase diagram with in grey the ``phases'' considered in the paper}
\end{figure}
Notice that, since only holons are involved and not full electrons, this is not a true phase transition as the one appearing in the DDW formalism \cite{Cha}. 

In PG,
    as a consequence of the $\pi$-flux, the holons are converted via Hofstadter
     mechanism into two species of Dirac fermions with small Fermi surface
     ($\epsilon_F \sim t \delta$) centered at the four nodes $\Bigl(\pm \frac {\pi}{ 2},
      \pm \frac {\pi}{ 2}\Bigr)$, whereas in SM they exhibit a large Fermi surface
 ($\epsilon_F \sim t (1-\delta)$), as expected from band structure calculations. A
 direct evidence
 of the small FS in PG might come from recent experiments on Shubnikov-de Haas
 oscillations \cite{doiron}.

Holons and spinons are gauge-invariantly coupled by a $U(1)$ slave-particle field, $A$,
 whose low energy effective action is  obtained upon integration of the matter fields.
  As a consequence of the finite FS of holons the transverse gauge propagator
 exhibits a Reizer singularity
\cite{Reizer} which  dominates  at large scales: for small $q,\omega, \omega/|\vec q|$

\begin{equation} \label{pi}
\langle A_\perp A_\perp \rangle (\omega,\vec q) \sim ( -\chi |\vec q |^2
+ i \kappa {\omega / |\vec q|})^{-1},
\end{equation}

where $A_\perp$ is the transverse component of $A$, $\chi$ is the diamagnetic
susceptibility and $\kappa$ the Landau damping. Both $\chi^{-1}$, $\kappa$ and
the holon mass $m_h$ are $\sim \delta$ in PG and $\sim 1-\delta$ in SM.
 The scalar component $A_0$ has a low energy propagator given
by
\begin{equation} \label{21} \langle A_0 A_0 \rangle (\omega, q) \sim (\kappa (1+
i \frac {\omega}{ |\vec q|})H(|\vec q|-|\omega|) +m^2_0)^{-1} \
\end{equation} where $m_0$ is a thermal mass generated by the
spinons and $H$ the Heaviside step function. \noindent In view of
the constant term in (\ref{21}) the interaction mediated by $A_0$
is short ranged, hence subleading at large distance w.r.t. the
interaction mediated by $A_\perp$. However, taking into account
the renormalization of the transverse contribution discussed
below, it gives the dominant contribution to $\gamma$ in PG  for
$T$ of the order of the holon Fermi
 temperature, which in this ``phase'' is rather small, of the order of a  hundred K at
 low dopings.
 In the gauge correlator the momenta extend up to an UV   cutoff $\Lambda \sim J$.

 As discussed in \cite{Lee, iof, Mar} at finite temperature $T$ the typical momentum scale
  of the transverse gauge fluctuations is given by the anomalous skin momentum
  $Q_0=(\kappa T/ \chi)^{1/3}$. The interaction of holon and spinon modifies the
  gauge propagator, inducing a cutoff of the infrared momentum singularity,
  for $|q| \lesssim Q_0$, replacing $\kappa |\vec q|^{-1}$ by the sum of the
  conductivity of the spinon-gauge and holon-gauge subsystems, denoted by $\sigma_s$
  and $\sigma_h$, respectively \cite{nota2}. The splitting in high and low momenta
  contributions is explicitly realized in the calculations with a sharp cutoff
  at $|\vec q|= \zeta Q_0$, where $\zeta \simeq 0.4$ \cite{nota1} and assuming
  a renormalization of $\kappa$ at high momenta accordingly.

To extract the entropy and the specific heat we start by computing the free energy $F$;
then $S= - \frac {\partial F}{\partial T}$ and $\gamma= - \frac {\partial^2 F}
{\partial T ^2}$, where we only differentiate w.r.t. the explicit dependence on $T$,
thus ensuring $S(T=0)=0$. Within our approach $F$ is the sum of four terms:
the contribution of free spinons, $F_s$, of free holons, $F_h$ and the fully
 renormalized contribution of transverse and scalar gauge fluctuations, $F_\perp, F_0$.
Since the spinons are massive, the $T$-dependence of $F_s$ is negligible for $T$ lower
 than the spinon gap, which we estimate of the order of few hundreds K. For the holons
  we have the standard result (for each holon species)
\begin{equation}
S_h \approx  c  m_h  T
\label{freh}
\end{equation}
where the phenomenological constant $c$ accounts also for the
eccentricity of the FS due to neglected $t',t''$ terms. A
comparison with \cite {yoshida} yields $c \approx 3$. We estimate
$F_\sharp,\sharp=\perp,0$ following Ref. \onlinecite{tsvelik}: if
we denote by $D_\sharp$ the fully renormalized retarded Green
function of the gauge field
\begin{equation}
F_\sharp \sim \int d \omega \coth (\omega/2 T) \int d^2 q \arctan(\frac{ {\rm Im}
D_\sharp(\omega, \vec q)}{ {\rm Re} D_\sharp(\omega,\vec q)}).
\end{equation}

We remark that in the calculation of $F_0$ in \cite{rice} a sophisticated ``ad hoc''
regularization was needed because the euclidean scalar correlator vanishes in the
limit of infinite frequency, thus making impossible a direct application of
$\zeta$-function regularization, since both spinons and holons are gapless in
the slave-boson approach. This problem does not arise here due to the constant
term in the scalar correlator caused by gapful spinons. The key result of \cite{rice}
that $F_0$ and $F_\perp$ have opposite sign is recovered here as a consequence of
the opposite relative sign of the real and immaginary part of $ D_\sharp$ for
scalar and transverse components, as follows from eqs. (\ref{pi}) and (\ref{21}).

The dominating contributions \cite{nota3} to entropy of transverse gauge fluctuations
turn out to be, up to logarithmic corrections,
\begin{equation}
  S_\perp \approx
   \left\{
   \begin{array}{ll} \label{fgt}
       Q_0^2 \sim T^{2/3} m_h^{4/3}
      & {\rm PG}  \\
      Q_0^2,  T \frac {\tilde\sigma}{\chi} H(Q_0^2- T \frac {\tilde\sigma}{\chi})
      &{\rm SM}
   \end{array}
   \right  .
\end{equation}
where $ \tilde\sigma= \sigma_h + \sigma_s \sim \tau_{imp} + \chi m_s^2 T^{-1}$
\cite{Mar1}. The $T^{2/3}$ behaviour is the standard one for 2D clean electrodynamics
\cite{tsvelik}. The second contribution in SM comes from ``small'' momenta and is
negligible in PG.
The leading scalar contribution is given by
\begin{equation} \label{fgs}
 S_0 \approx -\frac{\Lambda}{v_F} T + \frac{1}{v_F^2} T^2,
\end{equation}
where $v_F$ is the holon Fermi velocity. One can verify that
increasing $T$ first the transverse then the scalar contribution
dominates in PG, whereas in SM, in the temperature range
considered  the transverse contribution always dominates.

>From equations (\ref{freh}), (\ref{fgt}) and (\ref{fgs}) one can easily derive the
following consequences for the interpretation of experimental data within the
spin-charge gauge approach:

1) The approximately linear behaviour of $S$ is basically due to
the holons, although it is renormalized by gauge fluctuations. The
increase of the slope at low
 $\delta$ and its saturation at higher $T$ or $\delta$ are due to the transition
 from $m_h \sim \delta$ characteristic of PG to $m_h \sim 1-\delta$ characteristic
 of SM.

2) The negative intercept of entropy is due to the scalar gauge contribution (\ref{fgs})
in PG, negative in agreement with the general ideas discussed in the
introductory section.

3) The upturn $\Delta \gamma^{LT}$ in $\gamma$ (and the analogous more evident
in $S/T$ \cite{exp} ) is due to the contribution of transverse gauge fluctuations
in PG. Their contribution in SM yields the decrease above $T^*_\gamma$, which we
identify as the PG-SM crossover. These enhancement of entropy are due to the
presence of the gapless transverse gauge mode.

4) The approximately linear increase of $\gamma$ in $\delta$ and
in $T$, with almost $\delta$ independent slope, is due to the
second term in (\ref{fgs}) , presumably together with a pairing
contribution (see below). This behavior replaces the linear slope
due to the contribution of AF spin waves in the pristine material,
now removed by the spinon gap.

More concretely a comparison between theory and experiments is
summarized in figures 3 and 4, where in the inset the experimental
data for selected dopings are plotted only for the region of
parameters discussed above, where a comparison is
 meaningful. The theoretical curves have been obtained  with the same value of
 the parameters used in \cite{Mar,Mar1,mar} and
by substituting in $\tilde\sigma$ the expression for the conductivities derived
there. The value of $v_F$ in PG extracted from the slope of experimental $\gamma$
in the increasing range compared with the theoretical expression derived in
from (\ref{fgs}) turns out to be of the order of magnitude of the electron
Fermi velocity found experimentally in ARPES \cite{yoshida} and of the holon
Fermi velocity used in the calculation of transport properties in \cite{Mar, Mar1},
 although 2-3 times smaller. Presumably this is due to a further contribution of
 pairing , as in the preformed pair (see e.g. \cite{Lee, pre}) or fluctuating phase superconductor \cite{tes} approaches, which would yield a $T$-increasing density of states, not taken into
 account in the present simplified treatment. This increase should also account
 for the smooth transition from PG to SM discussed above in item 1). However a depletion mechanism of
 density of states (DOS),  alone seems to be unable to reproduce the experimental behavior of $\gamma$,
 because lowering $T$ the curves at different dopings should converge near $T=0$
  with increasing slope as $\delta$ increases (see e.g. \cite{Bi}, fig 12), whereas
  experimentally they are basically parallel at moderate temperatures with an
  upturn at low $T$ in non-superconducting samples.  The presence of two distinct effects appears consistent also with recent experimental data on specific heat in Bi$_2$Sr$_{2-x}$La$_x$CuO$_6$ \cite{hhwen}.
\begin{figure}
\centering
\includegraphics[scale=0.58]{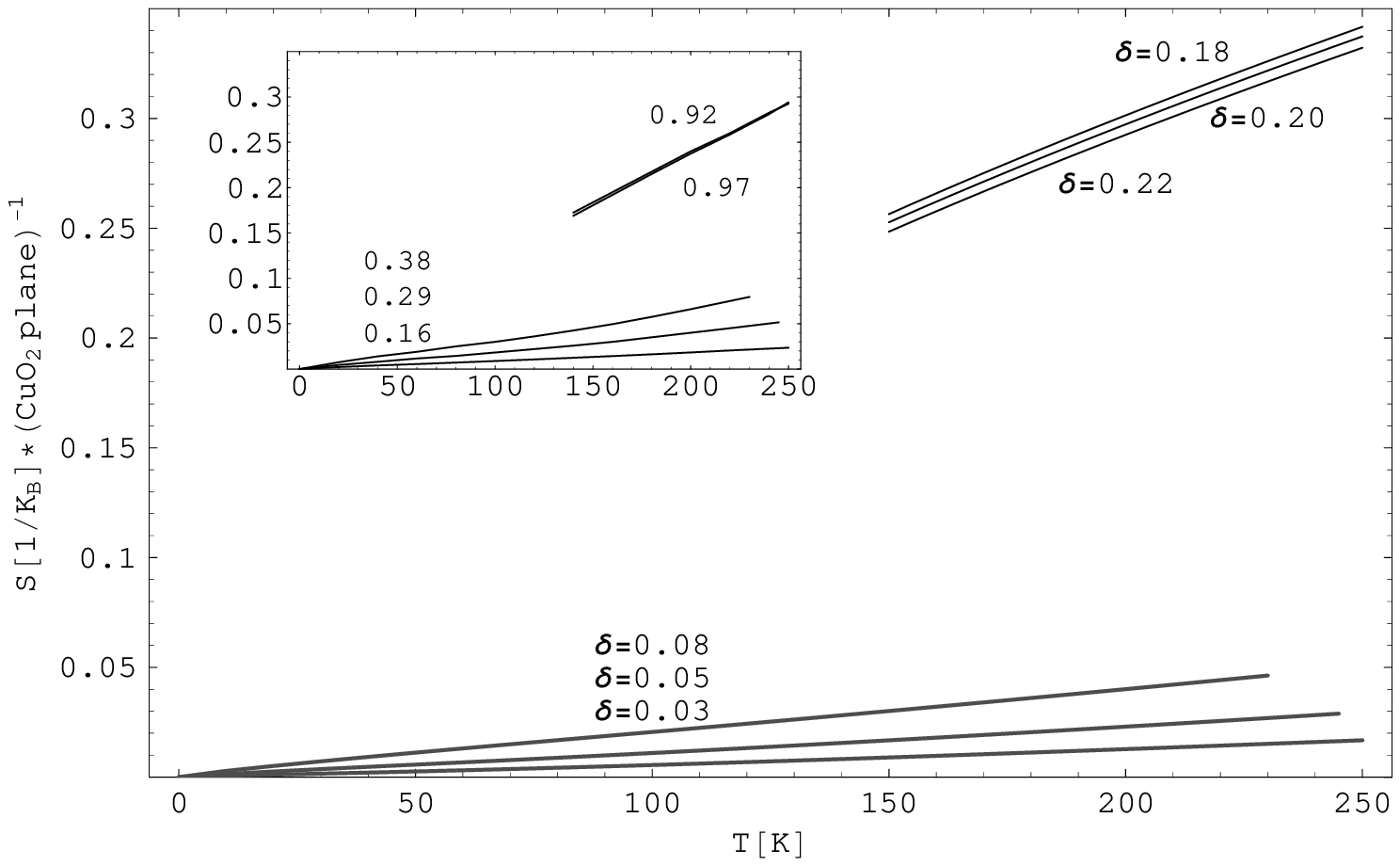}
\end{figure}

\begin{figure}
\centering
\includegraphics[scale=0.58]{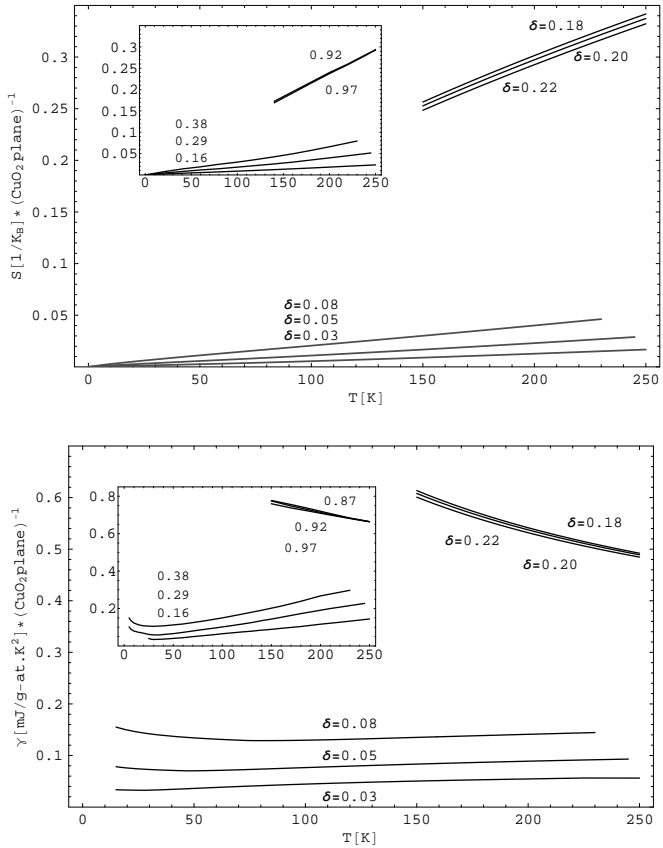}
\caption{The calculated temperature dependence for $S$ (above) and $\gamma$
(below) for selected dopings $\delta$ in both PG and SM ``phases'', in comparison
with experimental data for analogous in-plane doping concentrations in
YBa$_2$Cu$_3$O$_{6+x}$ ), taken from \cite{loram} (inset; labels show $x$).}
\end{figure}

As one can see, even in our simplified treatment a qualitative agreement of the
 behavior both in $T$ and $\delta$ is found deeply in PG and SM in the region of
 validity discussed above, reproducing the features discussed in the introductory
 section. The derived behavior of the entropy in SM is also consistent with the
 numerical data extracted from the $t$-$J$ model with the Lanczos method \cite{JP}
 or high temperature expansion \cite{rice} in the overlapping range of temperature.

\begin{figure}
\centering
\includegraphics[scale=0.58]{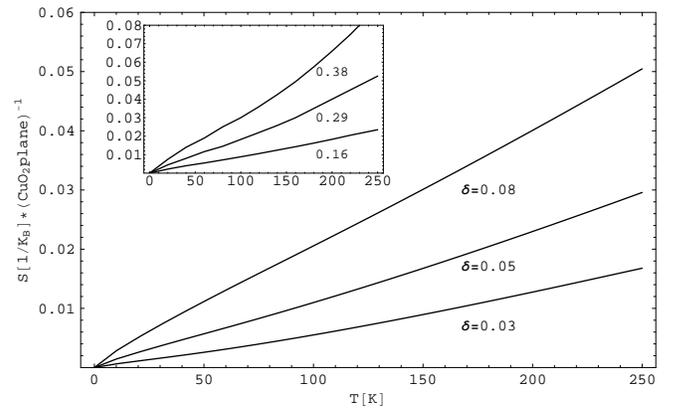}
\caption{Detail: Calculated $S$ in pseudogap ``phase'', in
comparison with experimental data on YBCO (inset) taken from
\cite{loram} with the same notations of
 Fig. 3}
\end{figure}

Summarizing, the reasonable agreement of doping-temperature dependence of experimental
 data and the theory discussed above suggests that one can interpret the low-temperature
 electronic specific heat in the normal state of hole-doped high Tc cuprates basically as due to
  fermionic holons, with small FS in the ``pseudogap'' and large FS in the ``strange
  metal phase'', but gauge fluctuations determine the variation in the $T$-dependence
   of the specific heat coefficient . Gauge fluctuations themselves reflect the changes
   in the spectrum of renormalized spin excitations and charge carriers.

In this approach the rather surprising negative intercept for the entropy is due to
the negative contribution of scalar gauge fluctuations. Since the gauge field is just
 a consequence of the Gutzwiller projection, this interpretation in terms of a
 ``constraint'' gauge field suggests that this peculiar feature is rooted in the
 no-double occupancy condition, hence a basic characteristic of the low-energy
 description of doped Mott insulators. The phenomenon appears in PG due to the
 ``effective'' rather small Fermi temperature, a consequence of the $\pi$-flux,
 characteristic of a 2D system, suppressing the positive transverse contribution.

Furthermore, in this approach the gauge field is the ``glue''
binding spinon and holon into an electron resonance \cite{Mar1,
mar}. The above interpretation then strengthens the idea of a composite
nature of the charge carriers in the cuprates, an idea which is
also strongly suggested by the metal-insulator crossover, as
discussed in detail in \cite{jcmp}.

{\bf Acknowledgments} It is  a pleasure for P.A.M. to thank Z.B. Su and L. Yu for
the joy of a long collaboration and G. Orso for pointing out a mistake in an earlier
 version of the manuscript. Very useful discussions with N. Hussey, T.M. Rice, F. Toigo, T. Xiang and H.H Wen are also gratefully acknowledged.


\begin{thebibliography}{9}
\bibitem{nota2}Notice that in view of the Ioffe-Larkin rule \cite{Iof} , holding in
 slave-particle gauge theories, the sum
$\tilde\sigma = \sigma_h + \sigma_s$ is not equal to the physical conductivity,
given by $
\sigma=(\sigma_h^{-1}+ \sigma_s^{-1})^{-1}.$
\bibitem{nota1} $\zeta$ is a constant O(1) fixed phenomenologically comparing with
 the experimental data
\bibitem{nota3} In the plot the full expression has been used.
\end{thebibliography}
\end{document}